# Quantized Majorana Conductance

Hao Zhang[1]*, Chun-Xiao Liu[2]*, Sasa Gazibegovic[3]*, Di Xu[1], John A. Logan[4], Guanzhong Wang[1], Nick van Loo[1], Jouri D.S. Bommer[1], Michiel W.A. de Moor[1], Diana Car[3], Roy L. M. Op het Veld[3], Petrus J. van Veldhoven[3], Sebastian Koelling[3], Marcel A. Verheijen[3,7], Mihir Pendharkar[5], Daniel J. Pennachio[4], Borzoyeh Shojaei[4,6], Joon Sue Lee[6], Chris J. Palmstrøm[4,5,6], Erik P.A.M. Bakkers[3], S. Das Sarma[2], Leo P. Kouwenhoven[1,8]

[1] QuTech and Kavli Institute of NanoScience, Delft University of Technology, 2600 GA Delft, the Netherlands

[2] Condensed Matter Theory Center and Joint Quantum Institute, Department of Physics, University of Maryland, College Park, Maryland 20742, USA

[3] Department of Applied Physics, Eindhoven University of Technology, 5600 MB Eindhoven, the Netherlands

[4] Materials Engineering, University of California Santa Barbara, Santa Barbara, CA, USA 93106

[5] Electrical and Computer Engineering, University of California Santa Barbara, Santa Barbara, CA, USA 93106

[6] California NanoSystems Institute, University of California Santa Materials, Santa Barbara, CA, USA 93106

[7] Philips Innovation Services Eindhoven, High Tech Campus 11, 5656AE Eindhoven, the Netherlands

[8] Microsoft Station Q Delft, 2600 GA Delft, The Netherlands

**Majorana zero-modes hold great promise for topological quantum computing. Tunnelling spectroscopy in electrical transport is the primary tool to identify the presence of Majorana zero-modes, for instance as a zero-bias peak (ZBP) in differential-conductance. The Majorana ZBP-height is predicted to be quantized at the universal conductance value of $2e^2/h$ *at zero temperature*. Interestingly, this quantization is a direct consequence of the famous Majorana symmetry, "particle equals antiparticle". The Majorana symmetry protects the quantization against disorder, interactions, and variations in the tunnel coupling. Previous experiments, however, have shown ZBPs much smaller than $2e^2/h$, with a recent observation of a peak-height close to $2e^2/h$. Here, we report a quantized conductance *plateau* at $2e^2/h$ in the zero-bias conductance measured in InSb semiconductor nanowires covered with an Al superconducting shell. Our ZBP-height remains constant despite changing parameters such as the magnetic field and tunnel coupling, i.e. a quantized conductance plateau. We distinguish this quantized Majorana peak from possible non-Majorana origins, by investigating its robustness on electric and magnetic fields as well as its temperature dependence. The observation of a quantized conductance plateau strongly supports the existence of non-Abelian Majorana zero-modes in the system, consequently paving the way for future braiding experiments.**

A semiconductor nanowire coupled to a superconductor can be tuned into a topological superconductor with two Majorana zero-modes localized at the wire ends[1-3]. Tunnelling into a Majorana mode will show a zero-energy state in the tunnelling density-of-states, i.e. a zero-bias peak (ZBP) in the differential conductance (d$I$/d$V$)[4,5]. This tunnelling process is a so-called Andreev reflection, where an incoming electron is reflected as a hole. Particle-hole symmetry dictates that the zero-energy tunnelling amplitudes of electrons and holes are equal, resulting in a perfect resonant transmission with a ZBP-height quantized at *$2e^2/h$*[6-8], irrespective of the precise tunnelling strength[9]. The Majorana-nature of this perfect Andreev reflection is a direct result of the well-known Majorana symmetry property "particle equals antiparticle"[10,11].

Such a predicted robust conductance quantization has not yet been observed[4,5,12-14]. Instead, most of the ZBPs have a height significantly less than $2e^2/h$. This discrepancy was first explained by thermal averaging[15-18]. This explanation, however, does not hold when the peak-width exceeds the thermal broadening (~$3.5k_BT$)[12,13]. In that case, other averaging mechanisms, such as dissipation[19], have been invoked. The main source of dissipation is a finite quasi-particle density-of-states within the superconducting gap, often referred to as a 'soft gap'. Substantial advances have been achieved in 'hardening' the gap by improving materials quality, eliminating disorder and interface roughness[20,21], and better control during device processing[22,23], all guided by a more detailed theoretical understanding[24]. We have recently solved all these dissipation and disorder issues[21], and here we report the resulting improvements in electrical transport leading to the so-far elusive quantization of the Majorana ZBP.

Fig.1a shows a micrograph of a fabricated device and schematics of the measurement set-up. An InSb nanowire (grey) is partially covered (two out of six facets) by a thin superconducting Al shell (green)[21]. The 'tunnel-gates' (coral red) are used to induce a tunnel barrier in the non-covered segment between the left electrical contact (yellow) and the Al shell. The right contact is used to drain the current to ground. The chemical potential in the segment proximitized by Al can be tuned by applying voltages to the two long 'super-gates' (purple).

Transport spectroscopy is shown in Fig.1b displaying d$I$/d$V$ as a function of voltage bias, $V$, and magnetic field, $B$ (aligned with the nanowire axis), while applying fixed voltages to the tunnel- and super-gates. As $B$ increases, two levels detach from the gap edge (at ~0.2 meV), merge at zero bias, and form a robust ZBP. This is consistent with the Majorana theory: a ZBP is formed after the Zeeman energy closes the trivial superconducting gap and re-opens a topological gap[2,3]. The gap re-opening is not visible in a measurement of the local density-of-states since the tunnel coupling to these bulk states is small[25]. Moreover, the finite length of the proximitized segment (~1.2 μm) results in discrete energy states, turning the trivial-to-topological phase transition into a smooth cross-over[26]. Fig.1c shows two line-cuts from Fig.1b extracted at 0 and 0.88 T. Importantly, the height of the ZBP reaches the quantized value of $2e^2/h$. The line-cut at zero-bias in the lower panel of Fig.1b shows that the ZBP-height remains close to $2e^2/h$ over a sizable range in *B*-field (0.75 - 0.92 T). Beyond this range, the height drops, most likely caused by a closure of the superconducting gap in the bulk Al shell.

We note that the sub-gap conductance at $B$ = 0 (black curve, left panel, Fig.1c) is not completely suppressed down to zero, reminiscent of a soft gap. In this case, this finite sub-gap conductance, however, does not reflect any finite sub-gap density-of-states in the proximitized wire. It arises from Andreev reflection (i.e. transport by dissipationless Cooper pairs) due to a high tunnelling transmission, which is evident from the above-gap conductance (d*I*/d*V* for *V* > 0.2 mV) being larger than $e^2/h$. Since this softness does not result from dissipation, the Majorana peak-height should still reach the quantized value[27]. In Extended Data Fig. 1, we show that this device tuned into a low transmission regime, where *dI/dV* does reflect the density-of-states, indeed displays a hard gap. For further understanding we use experimental parameters in a theoretical Majorana nanowire model[28] (see Methods for more information). Fig. 1d shows a simulation with two line-cuts in Fig. 1c (right panel). Besides the ZBP, other discrete sub-gap states are visible, which are due to the finite wire length. Such discrete lines are only faintly resolved in the experimental panels of Fig. 1b. Overall, we find good qualitative agreement between the experimental and simulation panels in Fig. 1b and 1d. We note that an exact quantitative agreement is not feasible since the precise experimental values for the parameters going into the theory (e.g. chemical potential, tunnel coupling, Zeeman splitting, spin-orbit coupling, etc.) are unknown for our hybrid wire-superconductor structure.

Next, we fix $B$ at 0.8 T and investigate the robustness of the quantized ZBP against variations in transmission by varying the voltage on the tunnel-gate. Fig. 2a shows d$I$/d$V$ while varying $V$ and tunnel-gate voltage. Fig. 2b shows that the ZBP-height remains close to the quantized value. Importantly, the above-gap conductance measured at |$V$| = 0.2 meV varies by more than 50% (Fig. 2c and 2d), implying that the transmission is changing significantly over this range while the ZBP remains quantized. Note that the minor conductance switches in Fig. 2a-c are due to unstable jumps of trapped charges in the surroundings.

Fig. 2d (red curves) shows several line-cuts of the quantized ZBP. The extracted height and width are plotted in Fig. 2e (upper panel) as a function of above-gap conductance $G_N$ = $T{\times}e^2/h$ where $T$ is the transmission probability for a spin-resolved channel. While the ZBP-width does change with $G_N$, the quantized height remains unaffected. Note that the ZBP-width ranges from ~50 to ~100 $\mu$eV, which is significantly wider than the thermal width ~6 $\mu$eV at 20 mK. The ZBP-width is thus broadened by tunnel coupling, instead of thermal broadening, i.e. fulfilling a necessary condition to observe a quantized Majorana peak. In Extended Data Fig. 2, we show that in the low transmission regime where thermal broadening dominates over tunnel broadening, the ZBP-height drops below $2e^2/h$[15-18]. We emphasize that the robustness of the ZBP quantization to a variation in the tunnel barrier is an important finding of our work.

A more negative tunnel-gate voltage (< -8 V) eventually splits the ZBP, which may be explained by an overlapping of the two localized Majorana wave-functions from the two wire ends. The tunnel-gate not only tunes the transmission of the barrier, but also influences the potential profile in the proximitized wire part near the tunnel barrier. A more negative gate voltage effectively pushes the nearby Majorana mode away, towards the remote Majorana on the other end of the wire, thus reducing the length of the effective topological wire. This leads to the wave-function overlap between the two Majoranas, causing the ZBP to split[16] (black curves in Fig. 2d). This splitting is also captured in our simulations shown in Fig. 2f, where we have checked that the splitting originates from Majorana wave-function overlap. Note that the simulated ZBP-height (red curve in middle panel in Fig. 2f) remains close to the $2e^2/h$ plateau over a large range, while the above-gap conductance (black curve in lower panel in Fig. 2f) changes substantially. Also, the height and width dependence in the simulation is in qualitative agreement with our experimental observation (Fig. 2e). To complete the comparison, we show in Fig. 2g the simulated line-cuts of several quantized ZBPs (red curves) and split peaks (black curves), consistent with the experimental data in Fig. 2d.

Pushing Majoranas toward each other is one mechanism for splitting. Another way is by changing the chemical potential through the transition from a topological to a trivial phase[2,3]— the topological quantum phase transition from the trivial to the topological phase can be equivalently caused by tuning either the Zeeman energy (i.e. the magnetic field) or the chemical potential. Splitting at the phase transition occurs since the Majorana wave-functions start to spread out over the entire wire length. For long wires the transition is abrupt, whereas in shorter wires a smooth transition is expected[26]. We investigate the dependence of the quantized ZBP on chemical potential by varying the voltage on the super-gate. Fig. 3a shows a nearly-quantized ZBP that remains non-split over a large range in the super-gate voltage. More positive voltage applied to the super-gates corresponds to a higher chemical potential, and eventually we find a ZBP-splitting (> -5 V) and consequently a suppression of the zero-bias conductance below the quantized value. Although the relation between the gate voltage and chemical potential is unknown in our devices, this splitting suggests a transition to the trivial phase caused by a tuning of the chemical potential induced by the changing super-gate voltage.

In a lower *B*-field and different gate settings (Fig. 3b), the splitting of the quantized ZBP shows oscillatory behaviour as a function of the super-gate voltage. The five line-cuts on the right panel highlight this back-and-forth behaviour between quantized and suppressed ZBPs. Remarkably, the ZBP-height comes back up to the quantized value and, importantly, does not cross through it.

We find similar behaviour in the theoretical simulations of Fig. 3c. In these simulations we have confirmed that for the chosen parameters, the Majorana wave-functions oscillate in their overlap, thus giving rise to the back-and-forth behaviour of quantized and split ZBPs[29]. In the experiment it may also be that non-homogeneity, possibly somewhere in the middle of the wire, causes overlap of Majorana wave-functions. Again, we note that the conversion from gate voltage to chemical potential is unknown, preventing a direct quantitative comparison between experiment and simulation.

To demonstrate the reproducibility of ZBP quantization, we show in Fig. 4a the quantized ZBP data from a second device. In this second device the length of the proximitized section is ~0.9 µm, which is ~0.3 µm shorter than the previous device. The quantized ZBP-plateau is indicated by the region between the two green dashed lines in Fig. 4b (red curve). This second device allows to transmit more than one channel through the tunnel barrier, which we deduce from the above-gap conductance value (Fig. 4b, lower panel, black curve) exceeding $e^2/h$ for tunnel-gate voltages higher than ~ -0.55 V. Correspondingly, the zero-bias conductance can now exceed $2e^2/h$ (Fig. 4b, middle panel) for such an open tunnel barrier[9]. We note that tunnelling through the second channel in the barrier region results in an additional background conductance, thus leading to the zero-bias conductance rising above $2e^2/h$. We find, however, from a rough estimate of this background contribution that the net ZBP-height (above background) never exceeds $2e^2/h$, consistent with Majorana theory[9].

We next fix the *B*-field and study temperature dependence. Fig. 4c shows a line-cut of this quantized ZBP from Fig. 4a. First, the base temperature trace in Fig. 4c (red data points) fits well to a Lorentzian line-shape with a 20 mK thermal broadening, expected for Majoranas as well as for any type of resonant transmission. The ZBP temperature dependence is shown in line traces in Fig. 4d and in colour scale in Fig. 4e (with the corresponding simulation in the lower panel of 4e). Fig. 4f shows the extracted ZBP-height and ZBP-width (i.e. full-width-half-maximum, FWHM) from both the experimental and simulational traces. At low temperatures, the ZBP-width (red data points) exceeds the thermal width defined as $3.5k_BT$ (blue line). In agreement with theory[30], the ZBP-height (black data points) reaches and saturates at $2e^2/h$ when the FWHM exceeds $3.5k_BT$. For higher temperatures, thermal averaging starts suppressing the ZBP-height below the quantized value. We note that the simulated data is calculated by a convolution of the derivative of the Fermi distribution function and the *dI/dV* trace at base temperature of 20 mK. This procedure of incorporating thermal effects holds if the temperature of the calculated *dI/dV* curve is significantly larger than base temperature (which can then be assumed to be the effective zero-temperature conductance value). Indeed, we find excellent agreement between experiment and simulation for *T* > 50 mK (Fig. 4f). See Extended Data Fig. 3 for detailed temperature dependence.

Recent theoretical work[28] has shown numerically for experimentally relevant parameters that ZBPs can also arise from local and non-topological Andreev bound states (ABS)[16,31-34]. These local ABS appear remarkably similar in tunnelling spectroscopy as the ZBPs arising from Majorana zero-modes. In a third device, we are able to find such non-topological states by fine-tuning gate voltages. Figure 5 shows the similarities and differences between ABS and Majorana ZBPs. First of all, Fig. 5a shows a ZBP in tunnelling spectroscopy versus *B*-field. At a particular *B*-field (0.7 T, red bar) the ZBP-height reaches $2e^2/h$. In this device, we next vary the chemical potential via a voltage applied to a back-gate, showing a fairly stable (non-split) ZBP in

Fig. 5b. In contrast, the ZBP is unstable against variations in tunnel-gate voltage. Fig. 5c shows that now the ZBP appears as level crossings instead of being rigidly bound to zero bias. The two different behaviours between back-gate and tunnel-gate are expected for ABSs that are localized near the tunnel barrier, as was modelled explicitly in Ref. 28 (see also Extended Data Fig. 5). Liu et al.[28] show that local ABSs can have near-zero energy, which in a *B*-field is remarkably robust against variations in chemical potential; in our experiment tuned by the back-gate. However, this is only the case for the tunnel-gate voltage fine-tuned to level crossing points at zero bias. The local tunnel-gate and the global back-gate thus have distinguishably different effects. For the Majorana case, instead of level crossing, the ZBP should remain non-split over sizable changes in tunnel-gate voltage[13], as shown in Fig. 2a and Fig. 4b.

The second fundamental difference is that the non-topological ABS ZBP-height is not expected to be robustly quantized at $2e^2/h$[9,28]. Fig. 5d and 5e show that the ZBP-height varies smoothly as a function of the back-gate voltage without any particular feature at $2e^2/h$. Also, the ZBP-height in Fig. 5a at $2e^2/h$ is just a tuned coincidence (see Extended Data Fig. 6). Note that the ZBP line-shape or temperature dependence does not discriminate between topological and non-topological cases. Both fit a Lorentzian line-shape as shown explicitly for the non-topological ABS in Fig. 5f. Thus, the temperature dependence alone cannot distinguish a Majorana origin from ABS[14,30,31]. Only a stable quantized tunnel-conductance plateau, robust against variations in all gate voltages and magnetic field strength, can uniquely identify a topological Majorana zero-mode as far as tunnelling spectroscopy is concerned.

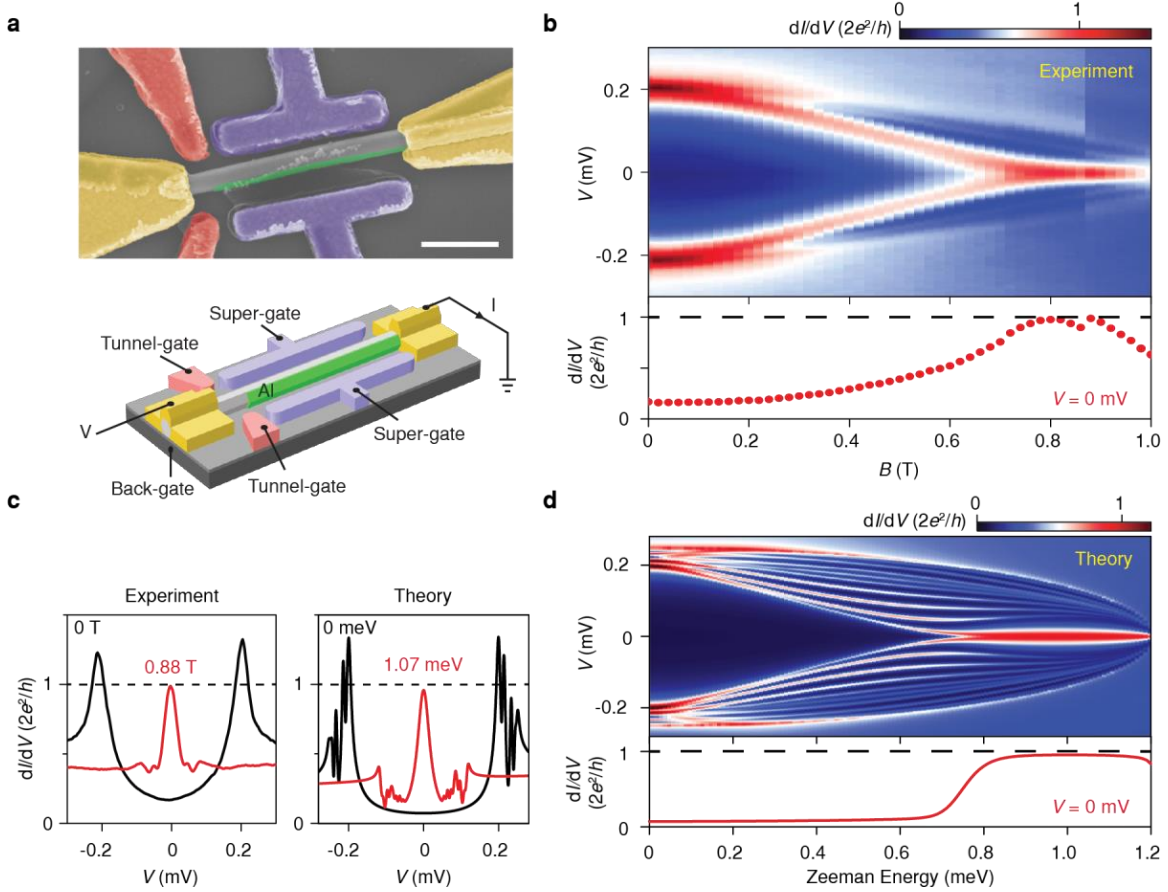

**Figure 1 | Quantized Majorana zero-bias peak. a,** False-colour scanning electron micrograph of device A (upper panel) and its schematics (lower panel). Side gates and contacts are Cr/Au (10 nm/100 nm). The Al shell thickness is ~10 nm. The substrate is p-doped Si, acting as a global back-gate, covered by 285 nm $SiO_2$. The two tunnel-gates are shorted externally as well as the two super-gates. The scale bar is 500 nm. **b,** Magnetic field dependence of the quantized ZBP in device A with the zero-bias line-cut in the lower panel. Magnetic field direction is aligned with the nanowire axis for all measurements. Super- (tunnel-) gate voltage is fixed at -6.5 V (-7.7 V), while the back-gate is kept grounded. Temperature is 20 mK unless specified. **c,** Comparison between experiment and theory. Left (right) panel shows the vertical line-cuts from **b** (**d**) at 0 T and 0.88 T (1.07 meV). **d,** Majorana simulation of device A, assuming chemical potential μ = 0.3 meV, tunnel barrier length ($L_{TG}$ = 10 nm), with height $E_{TG}$ = 8 meV, and the superconductor-semiconductor coupling is 0.6 meV. See Methods for further information. A small dissipation broadening term (~30 mK) is introduced for all simulations to account for the averaging effect from finite temperature and small lock-in excitation voltage (8 μV)

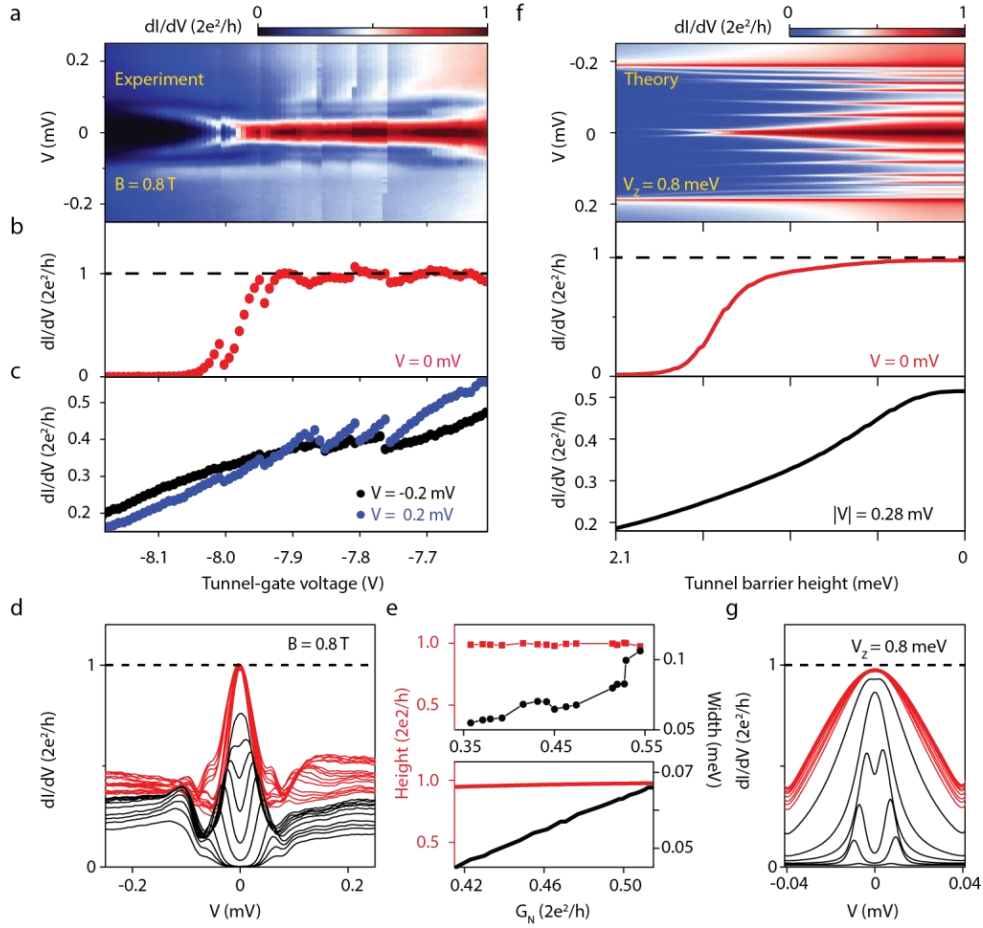

**Figure 2 | Quantized Majorana conductance plateau. a,** Tunnel-gate dependence of the quantized ZBP at $B = 0.8$ T. Super- (back-) gate voltage is fixed at -6.5 V (0 V). **b,c,** Horizontal line-cuts from **a**, showing zero-bias conductance and above-gap conductance, respectively. The zero-bias conductance shows a quantized plateau. **d,** Several vertical line-cuts from **a**, showing ZBPs with quantized height (red curves). For the black curves the zero-bias conductance drops below the quantized value due to peak splitting. **e, (upper panel)** The ZBP-height (red dots) and width (black dots) extracted from **d** (red curves), as a function of above-gap conductance ($G_N$). The width is defined by the bias voltage at which $dI/dV = e^2/h$. **(lower panel)** ZBP-height and width extracted from several simulation curves in **f**. **f,** Majorana simulation of the tunnel-gate dependence. We set the Zeeman field $V_Z = 0.8$ meV, chemical potential $\mu = 0.6$ meV, such that the nanowire is in the topological regime. From left to right, the barrier width decreases linearly from 175 to 0 nm, as the barrier height decreases from 2.1 meV to 0. **g,** Vertical line-cuts from **f** show the quantized ZBP (red) and split-peaks (black).

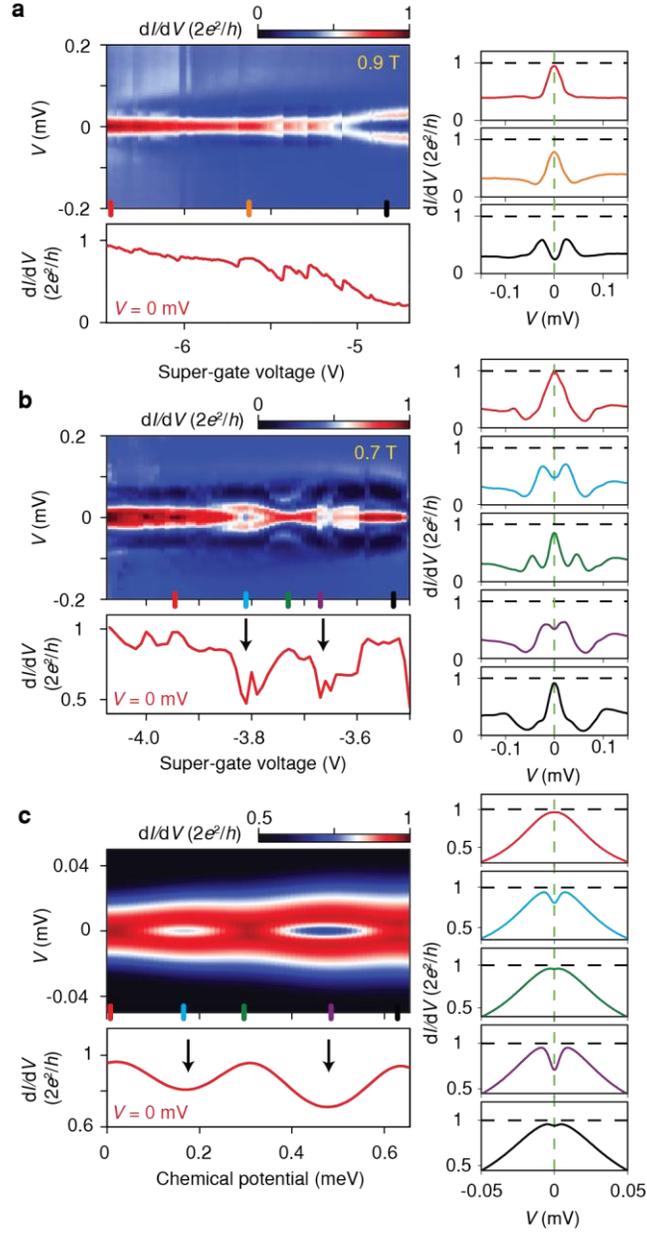

**Figure 3 | Majorana peak splitting. a,** Super-gate dependence of the quantized ZBP in device A at 0.9 T. As the super-gate increases the chemical potential, the ZBP-height is nearly quantized before it splits. The tunnel-gate voltage is adjusted simultaneously when sweeping the super-gate voltage, to compensate for the cross coupling and keep the transmission roughly constant. Lower panel shows the zero-bias line-cut, and the right panels show vertical line-cuts at gate voltages indicated by the corresponding colour bars. Switches in the colour maps are due to charge jumps in the gate dielectric. **b,** Oscillatory behaviour of the ZBP splitting, where the two black arrows point at the peak splitting regions. **c,** Simulation also shows oscillatory splitting as a function of chemical potential. The Zeeman field is fixed at $V_Z = 1$ meV.

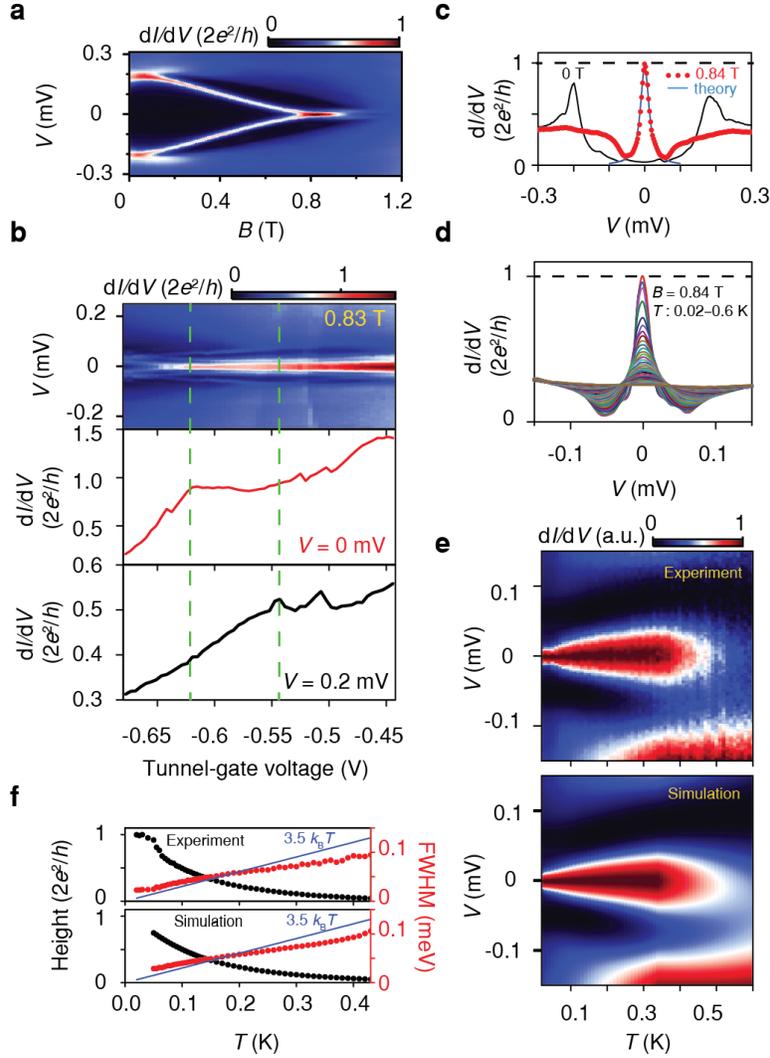

**Figure 4 | Quantized Majorana plateau reproduced, and temperature dependence. a**, Magnetic field dependence of the quantized ZBP in device B. **b,** Tunnel-gate dependence of the ZBP at 0.83 T. The two lower panels are the horizontal line-cuts at bias voltage, *V*, of 0 and 0.2 mV. The two dashed green lines indicate the plateau region of the zero-bias conductance. **c,** Vertical line-cuts from **a** at 0 T and 0.84 T. The blue line is a Lorentzian fit with a tunnel coupling $\Gamma = 13.7$ μeV and temperature of 20 mK. **d,** Temperature dependence of this quantized ZBP while the temperature increases from 20 mK to 600 mK in steps of 10 mK. **e,** Colour plot of the temperature dependence in the upper panel with the simulation in the lower panel. At each temperature the conductance is renormalized by setting the minimum to 0 and maximum to 1, for clarity. **f,** Extracted ZBP-height and FWHM as a function of temperature from **e**. Upper panel is the experiment while the lower panel is the simulation with no fitting parameters.

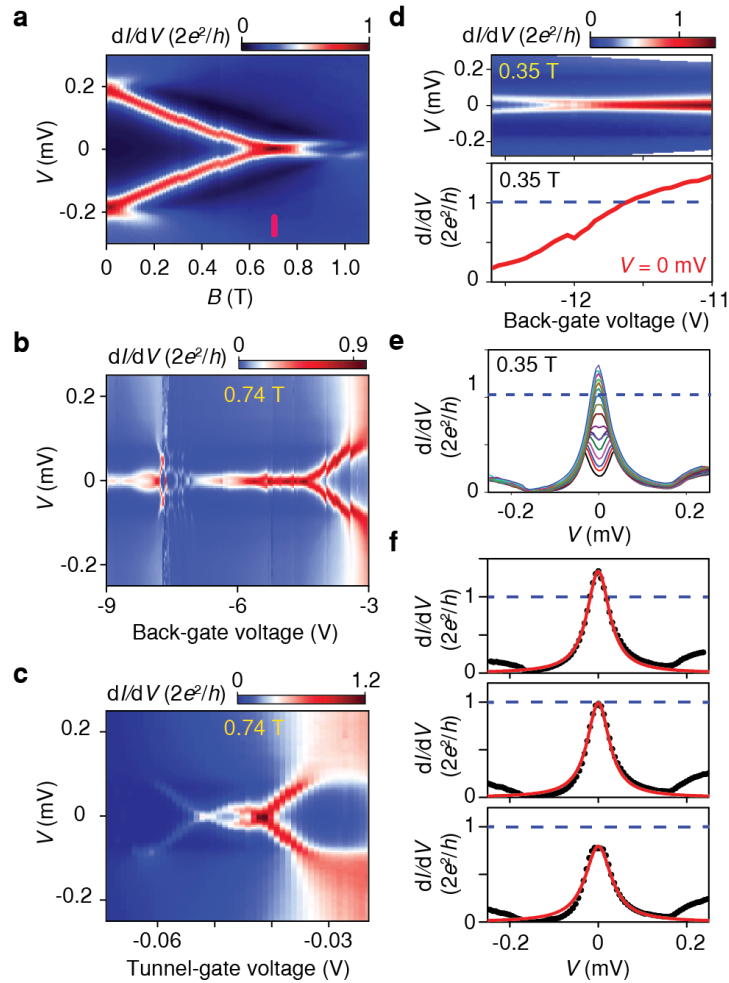

**Figure 5 | Trivial zero-bias peaks from Andreev bound states. a**, Magnetic field dependence of a trivial ZBP in device C. The peak-height reaches $2e^2/h$ at 0.7 T (red bar). **b,** Back-gate dependence of this ZBP, where the peak remains non-split for a sizable gate voltage range. The peak-height varies, and is generally below $2e^2/h$ **c,** Tunnel-gate dependence of this ZBP, which is a result of level crossing. **d,** Back-gate dependence of the ZBP at 0.35 T, with the lower panel showing the zero-bias line-cut. **e,** Vertical line-cuts from **d**. **f,** Lorentzian fit (red curve) of three ZBP curves (black dots) taken from **e**, assuming a temperature broadening of 20 mK.

*These authors contributed equally to this work

**Acknowledgments:** We acknowledge stimulating discussions with Michael Wimmer and Önder Gül. This work has been supported by the European Research Council (ERC), the Dutch Organization for Scientific Research (NWO), Office of Naval Research (ONR), Laboratory for Physical Sciences (LPS) and Microsoft Corporation Station-Q.

**Author Contributions:** The teams in Eindhoven and Santa Barbara have grown the nanowires with epitaxial Al and performed the nanowire deposition. The team in Delft fabricated the devices, performed electrical measurements, and analysed the experimental data. The Maryland team performed the numerical simulations. The manuscript was written by HZ and LPK with comments from all authors.

**Author Information:** The authors declare no competing financial interests. Correspondence should be addressed to H.Zhang-3@tudelft.nl (HZ); Leo.Kouwenhoven@microsoft.com (LPK)

**Method**

**Theory model.** We use the theoretical model from reference 28 to perform numerical simulations with experimentally relevant parameters, such as the effective mass $m^* = 0.015\ m_e$, the spin-orbit coupling $\alpha = 0.5$ eVÅ, the chemical potential of the normal metal lead $\mu_{lead} = 25$ meV, the Landé $g$-factor $g = 20$ such that the Zeeman energy $V_Z$ [meV] = 1.2 B [T], and the length of the nanowire $L = 1.0\ \mu$m. Note that the collapse of the bulk Al superconducting gap is included explicitly in the theory to be consistent with the experimental situation where the bulk gap collapses ~ 1T.

**Lorentzian fit.** We fit our ZBP line-shape with the Lorentzian formula: $G(V) = \frac{2e^2}{h} \times \frac{\Gamma^2}{\Gamma^2 + (eV)^2}$, where $\Gamma$ defines the tunnel coupling and FWHM of the peak, i.e. $2\Gamma$. Then we do convolution integration with the derivative of the Fermi distribution function (at 20 mK) to fit our ZBP shape. Since the FWHM of our ZBP is much larger than the thermal width, we took $\Gamma$ to be roughly equal to half of the FWHM for all the fittings in Fig. 4c and Fig. 5f.

# Extended Data: Quantized Majorana Conductance

Hao Zhang[1]*, Chun-Xiao Liu[2]*, Sasa Gazibegovic[3]*, Di Xu[1], John A. Logan[4], Guanzhong Wang[1], Nick van Loo[1], Jouri D.S. Bommer[1], Michiel W.A. de Moor[1], Diana Car[3], Roy L. M. Op het Veld[3], Petrus J. van Veldhoven[3], Sebastian Koelling[3], Marcel A. Verheijen[3,7], Mihir Pendharkar[5], Daniel J. Pennachio[4], Borzoyeh Shojaei[4,6], Joon Sue Lee[6], Chris J. Palmstrøm[4,5,6], Erik P.A.M. Bakkers[3], S. Das Sarma[2], Leo P. Kouwenhoven[1,8]

[1] QuTech and Kavli Institute of NanoScience, Delft University of Technology, 2600 GA Delft, the Netherlands

[2] Condensed Matter Theory Center and Joint Quantum Institute, Department of Physics, University of Maryland, College Park, Maryland 20742, USA

[3] Department of Applied Physics, Eindhoven University of Technology, 5600 MB Eindhoven, the Netherlands

[4] Materials Engineering, University of California Santa Barbara, Santa Barbara, CA, USA 93106

[5] Electrical and Computer Engineering, University of California Santa Barbara, Santa Barbara, CA, USA 93106

[6] California NanoSystems Institute, University of California Santa Materials, Santa Barbara, CA, USA 93106

[7] Philips Innovation Services Eindhoven, High Tech Campus 11, 5656AE Eindhoven, the Netherlands

[8] Microsoft Station Q Delft, 2600 GA Delft, The Netherlands

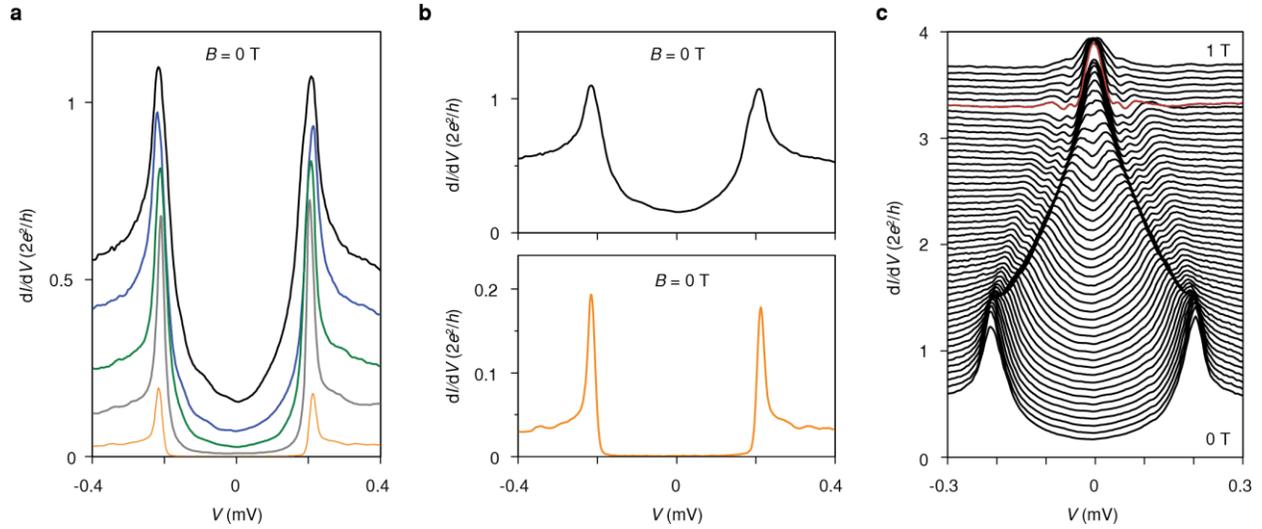

**Extended Data Figure 1 | Apparent "soft gap" due to large Andreev reflection. a,** d$I$/d$V$ of the device in Fig. 1-3 (device A) as a function of bias voltage at zero magnetic field. The tunnel-gate voltage is tuned to more negative from the top curve to the bottom curve. The transmission probability of the tunnel barrier is tuned from large (black curve) to small (orange curve). In the low transmission regime (orange curve), where the above-gap conductance (~ 0.03 × $2e^2/h$) is much less than $2e^2/h$, d$I$/d$V$ is proportional to the density of states in the proximitized wire part, resolving a hard superconducting gap. In the high transmission regime (black curve), where the above-gap conductance is comparable with $2e^2/h$, the finite sub-gap conductance is due to large Andreev reflection. This "soft gap" is *not* from dissipation, and does *not* affect the quantized ZBP-height as shown in **c**. **b,** Re-plot of the two curves from **a**. **c,** Waterfall plot of Fig. 1b, showing all the individual curves from 0 T to 1 T in steps of 0.02 T. The curves are offset vertically by 0.066 × $2e^2/h$ for clarity. The curve at 0 T and the red curve at 0.88 T correspond to the curves in Fig. 1c (left panel).

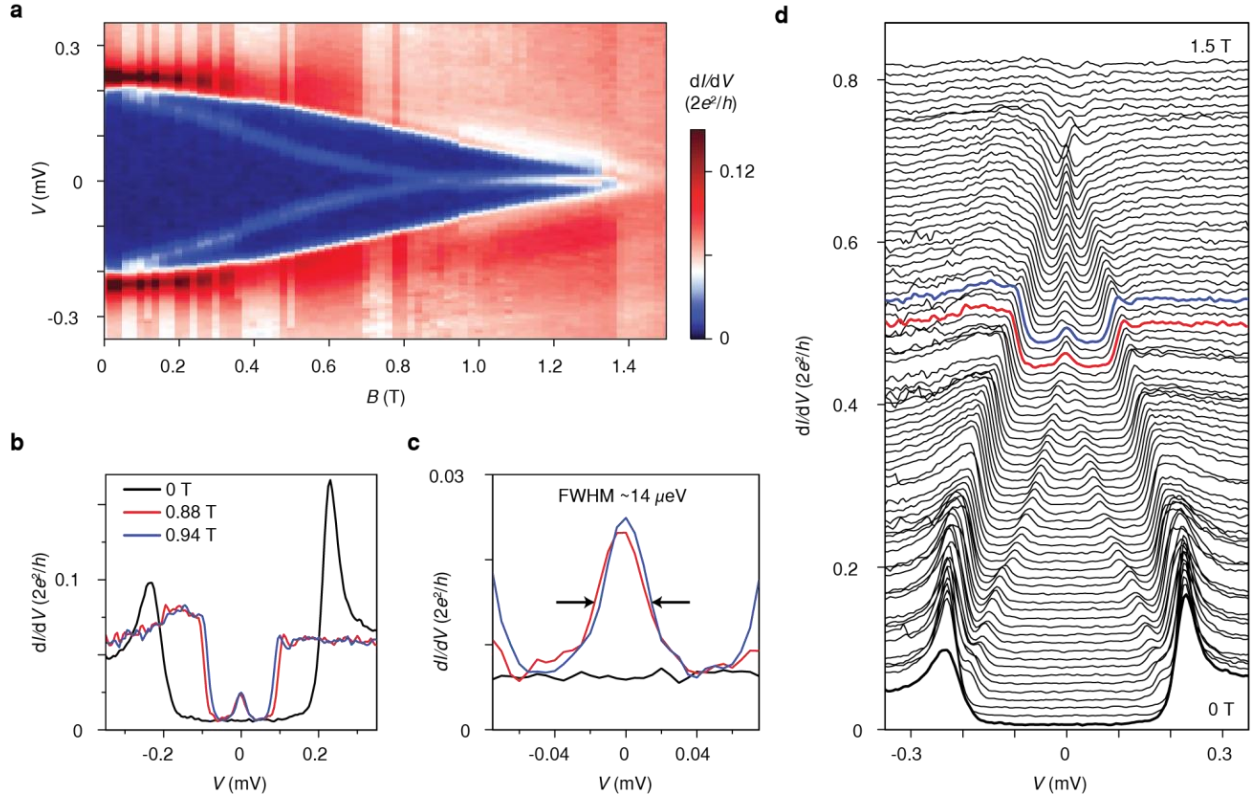

**Extended Data Figure 2 | Thermal-broadened ZBP in low transmission regime. a,** d$I$/d$V$ of device D, as a function of $B$, showing a stable ZBP. **b,** Vertical line-cuts at 0 T, 0.88 T and 0.94 T. At $B$ = 0 T, the above-gap conductance (~0.05×2$e^2$/h) is much less than $2e^2/h$, which means the device is in the low transmission regime, and thus shows a hard gap. The tiny sub-gap conductance is due to the small Andreev reflection and the noise background of the measurement equipment. The low transmission leads to a narrow ZBP-width, which is negligible compared to the thermal width of 3.5$k_B$T. Thus thermal averaging suppresses the ZBP-height below the quantized value. The sub-gap conductance at finite $B$ (e.g. 0.88 T or 0.94 T), where the ZBP appears, is the same as the sub-gap conductance at zero field, indicating that the gap remains hard at high magnetic field where the Majorana state is present. **c,** The zoom-in curves show that the FWHM of the ZBP ~14 μeV, which is consistent with the combined effect of the thermal broadening (3.5 $k_B$T ~ 6 μeV at 20 mK), the lock-in bias voltage excitation (5 μeV) and broadening from tunnelling. This shows that the thermal broadening indeed dominates over tunnel broadening. **d,** Waterfall plot of **a** with vertical offset of 0.01 × 2$e^2$/h for clarity.

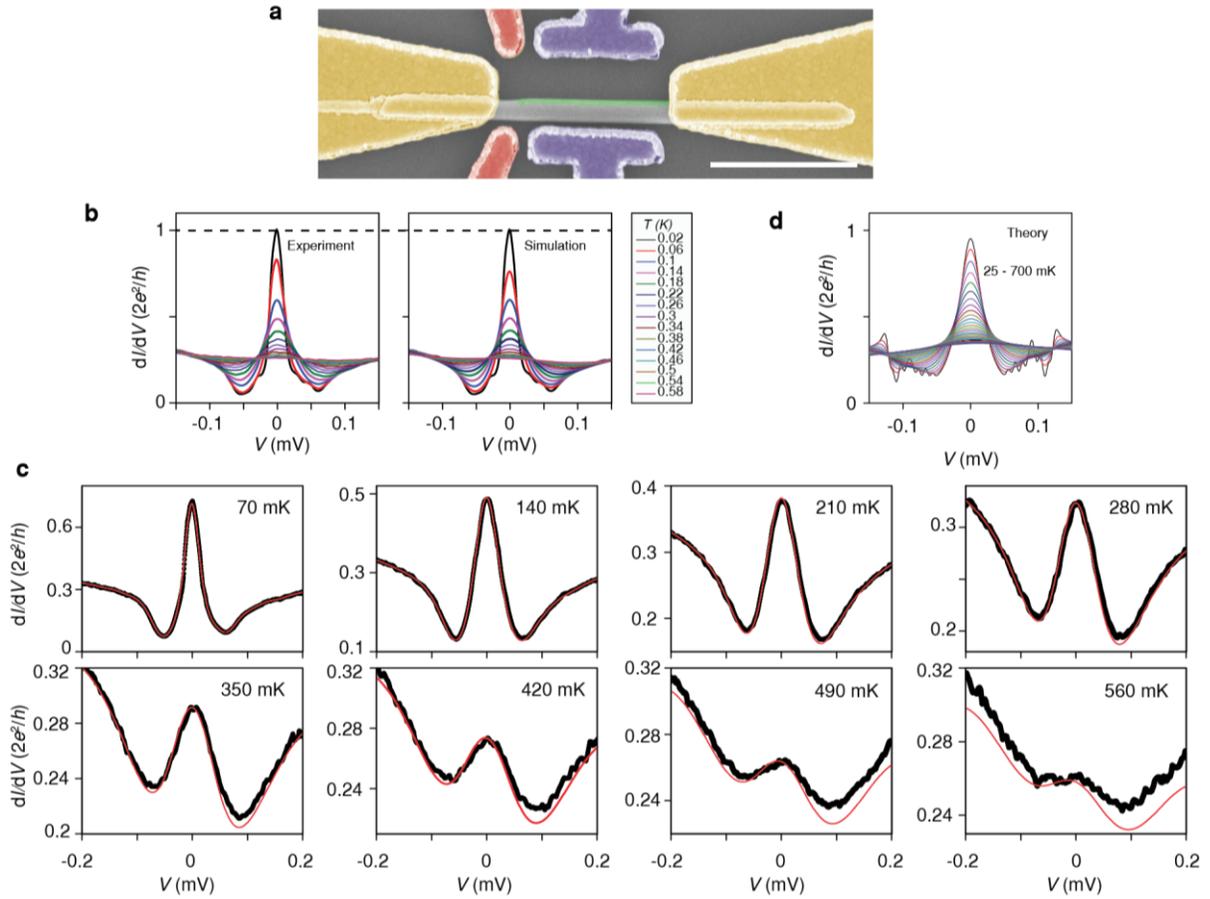

**Extended Data Figure 3 | Simulation of temperature dependence on the quantized ZBP. a,** False-colour scanning electron micrograph of device B with data shown in Fig. 4. Scale bar is 1 μm. The length of the Al section is ~0.9 μm. We calculate the d$I$/d$V$ curve at high temperature by convolution of the derivative of the Fermi distribution function with the d$I$/d$V$ curve at base temperature of 20 mK: $dI/dV = G(V,T) = \int_{-\infty}^{\infty} d\epsilon\, G(\epsilon, 0) \frac{df(eV-\epsilon,T)}{d\epsilon}$, where $T$ is temperature, $V$ is bias voltage, and $f(E,T)$ is the Fermi distribution function. Since we use the d$I$/d$V$ curve at 20 mK as the zero temperature data, our model only works for $T$ sufficiently larger than 20 mK, i.e. $T > 50$ mK. **b,** Comparison between the experimental data (left, taken from Fig. 4d) and theory simulations, for different temperatures. **c,** Several typical curves at different temperatures, black traces are the experimental data while the red curves are the theory simulations with no fitting parameters. The perfect agreement between simulation and experiment indicates that thermal averaging effect is the dominating effect that smears out the ZBP at high temperature. **d,** Temperature dependence the ZBP taken from our theory model: Fig. 1c (right panel). The temperature varies from 25 mK to 700 mK in steps of 23 mK.

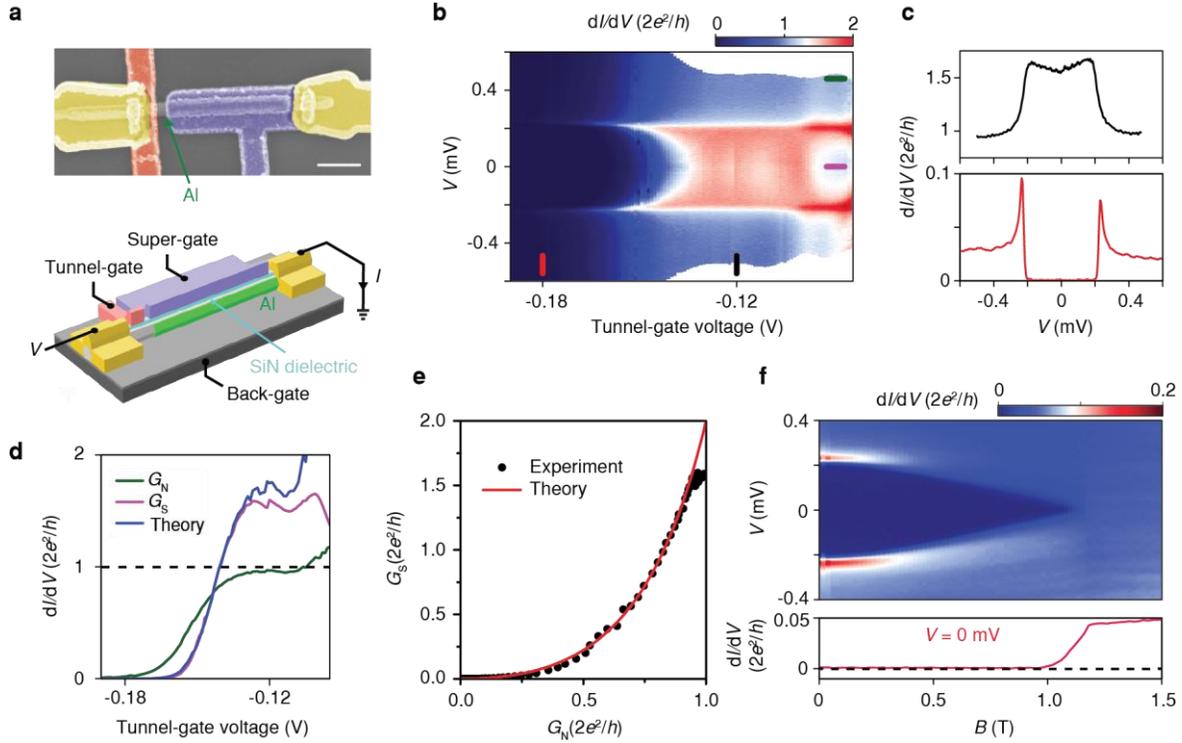

**Extended Data Figure 4 | Perfect ballistic Andreev transport in InSb-Al nanowires. a,** False-colour scanning electron micrograph of the device in Fig. 5 (device C). Scale bar is 500 nm. Electrical contacts and top gates are Cr/Au. Lower panel shows the device schematic and measurement set-up. The two top-gates (tunnel-gate and super-gate) are separated from the nanowire by 30 nm thick SiN dielectric. The global back gate is p-doped Si covered by 285 nm thick $SiO_2$ dielectric. **b,** $dI/dV$, as a function of bias voltage ($V$) and tunnel-gate voltage at zero field. No localization effect (conductance resonances or quantum dot induced Coulomb blockade) is observed. **c,** Vertical line-cuts from **b** at tunnel-gate voltage of -0.18 V (lower panel) and -0.12 V (upper panel), showing a hard superconducting gap in the low transmission regime (lower panel) and strong Andreev enhancement in the open regime (upper panel). **d,** Horizontal line-cuts from **c** for $V = 0$ mV (pink, sub-gap conductance, $G_S$) and $V = 0.45$ mV (green, above-gap conductance, $G_N$). The blue curve is the calculated sub-gap conductance using $G_S = 4e^2/h \times T^2/(2-T)^2$, where the transmission $T$ is extracted from the above-gap conductance: $G_N = 2e^2/h \times T$. **e,** $G_S$ as a function of $G_N$ (black dots) and the theory prediction (red curve): $G_S = 2 \times G_N^2/(2-G_N)^2$, with $G_S$ and $G_N$ in unit of $2e^2/h$. Both panel **d** and **e** show perfect agreement between theory and experiment. This indicates that the sub-gap conductance is indeed dominated by the Andreev reflection, i.e. without contributions from sub-gap states. **f,** Magnetic field dependence of the hard gap. Lower panel shows the zero-bias line-cut. The gap remains hard up to 1 Telsa, where the bulk superconducting gap closes.

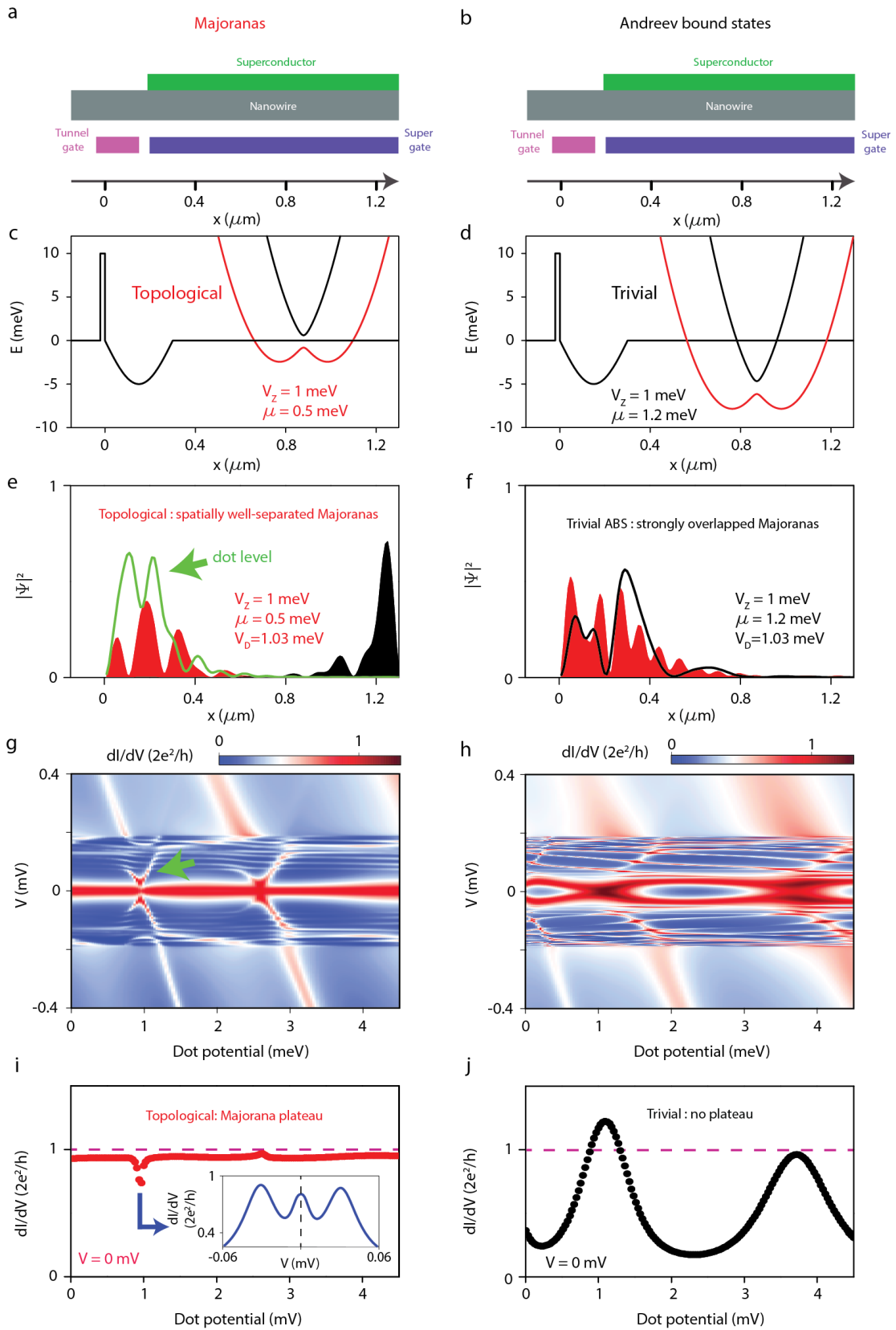
23

**Extended Data Figure 5 | Majoranas versus trivial Andreev bound states. a&b,** Schematics of a Majorana nanowire device. The only difference between the left column (Majorana) and right column (ABS) is the chemical potential, as shown in **c&d**: potential profile in the device. The tunnel barrier height is 10 meV and the width is 10 nm. The dot potential shape is $E(x) = -V_D \sin(\pi x/l_{dot})$, for x between 0 and 0.3 µm, where the length of the dot ($l_{dot}$) is 0.3 µm, $V_D$ is the dot depth which can be tuned by the nearby gate, i.e. the tunnel-gate. The rest of the flat nanowire segment is 1 µm long. We assume a pairing potential $\Delta$ = 0.2 meV, with a spin-orbit coupling of 0.5 eVÅ. We set the Zeeman energy to be 1 meV, thus the chemical potential of 0.5 meV (left) corresponds to the topological regime, and 1.2 meV (right) corresponds to the trivial regime, based on the topological condition: $V_Z > \sqrt{\mu^2 + \Delta^2}$. **e&f,** Spatial distribution of the Majorana and ABS wave-functions in the topological and trivial regime. In the topological regime, two spatially well separated Majoranas (red and black) are localized at the two ends of the topological section. In the trivial regime, the Andreev bound state, which can be considered as two strongly overlapped Majoranas (red and black), is localized near the tunnel barrier. **g&h,** The Majorana ZBP remains non-split against the change of dot potential, regardless of the energy of the dot level. The green arrow indicates one bound state in the dot, whose wave-function is shown in **e** (green curve). When this dot level moves down, it is repelled from zero energy, where the Majorana ZBP remains bound to zero (inset of **i**). On the contrary, the ABS-induced ZBP is not robust at all and only show up at the crossing points of two Andreev levels. This is because the tunnel-gate tunes the dot potential, which therefore affects the energy of the localized ABS near the tunnel barrier. **i&j,** The Majorana ZBP-height shows a quantized plateau at *2e²/h* by tuning the dot potential with tunnel-gate. The ZBP-height drops from the quantized value (inset) when the ABS-dot level moves towards zero, which effectively squeezes the ZBP-width such that the thermal averaging effect starts to dominate. The ABS zero-bias conductance does not show a plateau, but instead varies between 0 and 4*e²/h*.

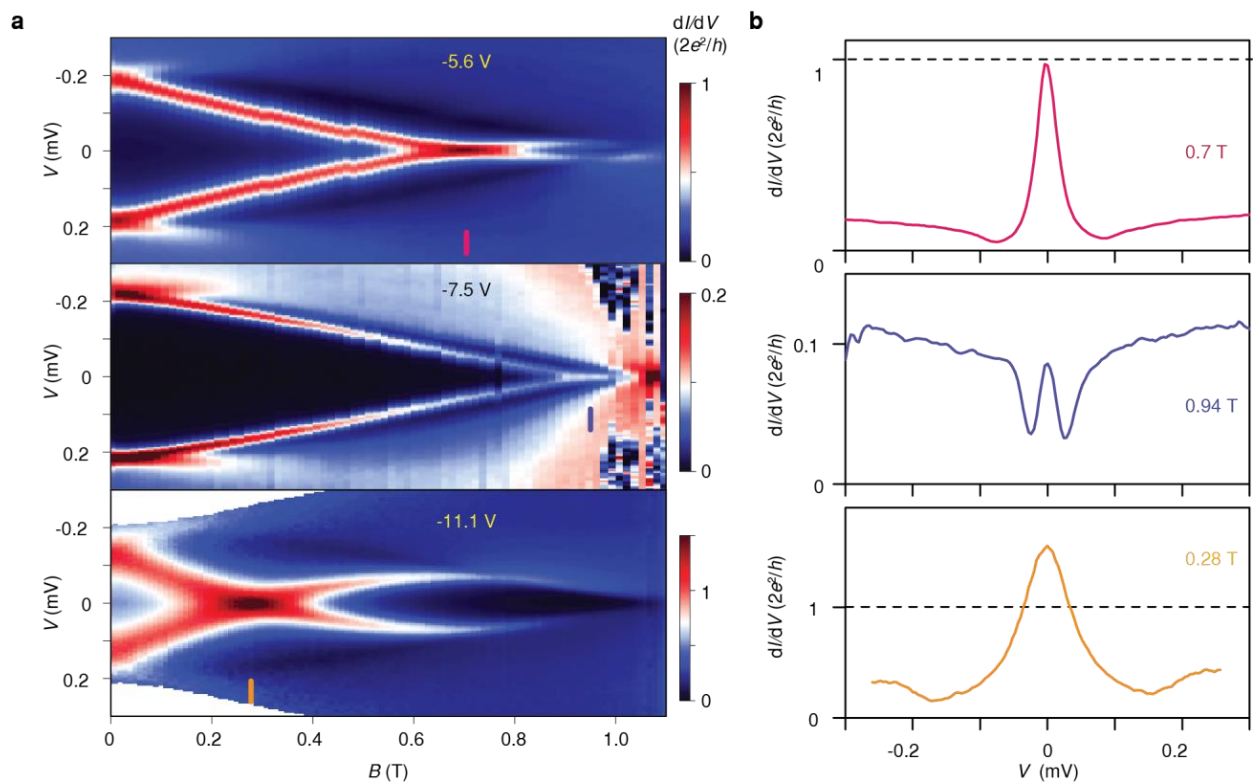

**Extended Data Figure 6 | Magnetic field dependence of trivial Andreev bound states a,** Top panel is a re-plot of the trivial ABS data in Fig. 5a. Middle and bottom panels are the ZBP data at different back-gate voltages (labelled in the panels). **b,** Line-cuts of the ZBP data from **a**. The ZBP-height varies with back-gate voltages, and can exceed $2e^2/h$. The ZBP-height at $2e^2/h$ here is just a tuned coincidence.

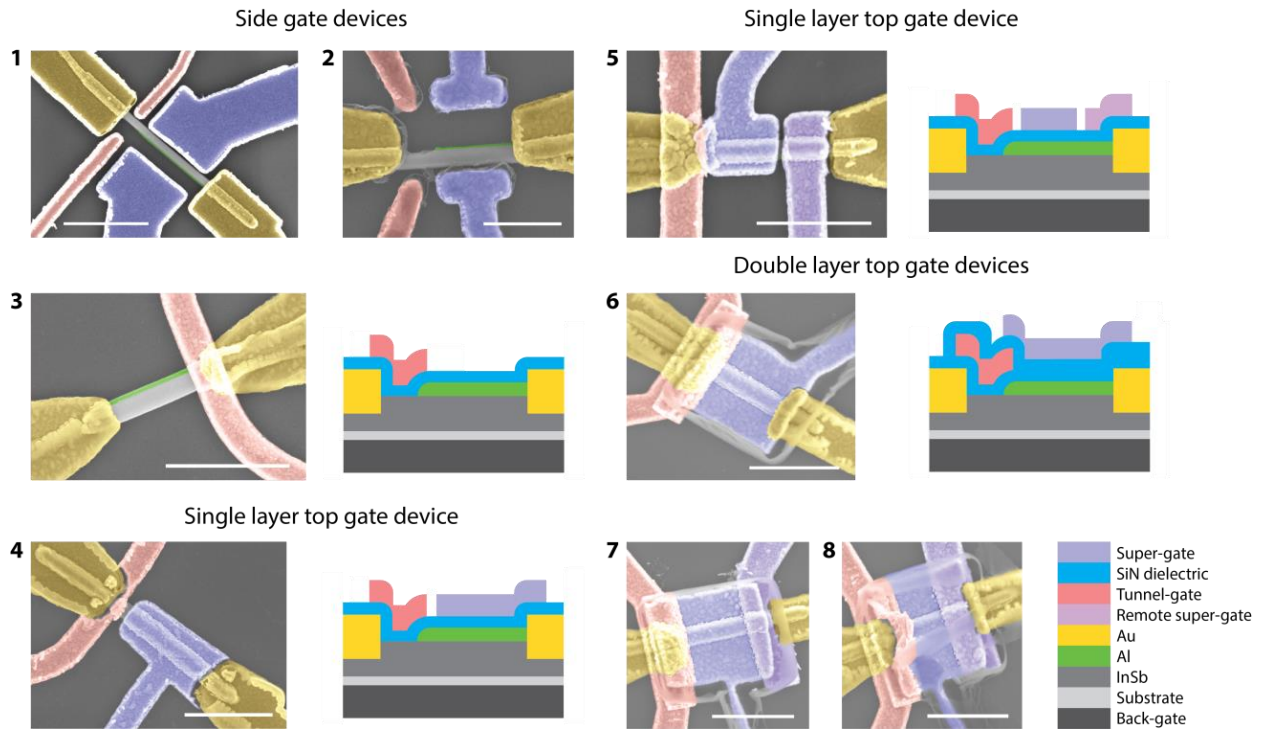

**Extended Data Figure 7 | Specifics of devices.** We fabricated and tested many (>60) devices out of which we selected *11* devices, which did show good basic transport with all gates being fully functional. These were used for extensive measurements. While most of these devices show ZBPs after tuning gate voltages and magnetic field, only *2* devices (presented in the manuscript, Figs. 1,2,3 for device A and Fig.4 for device B) show a *quantized* ZBP-plateau. All other devices show trivial ZBPs similar to Fig.5 (from device C). The SEM images of device A, B & C are shown in Fig. 1a, Extended Data Fig. 3a and Extended Data Fig. 4a, respectively.

Here we show the SEM images of all other *8* devices, which we have explored extensively, but without finding a quantized ZBP-plateau. Dev.1&2 are side gate devices. Dev.3 has a top tunnel-gate separated from the nanowire by 30 nm thick SiN dielectric, and a global back gate separated by 285 nm thick $SiO_2$. Dev. 4&5 have tunnel-gate and super-gate on top separated from the nanowire by 30 nm thick SiN dielectric. Dev. 6-8 have two layers of top gate. The bottom layer has a tunnel-gate separated by 30 nm thick SiN dielectric while the top layer has super-gates separated by 30 nm thick SiN from the bottom layer. The scale bar is 1 µm for all devices except for device 2, which is 500 nm.

It would be very informative to perform Schrodinger-Poisson calculations on these different device geometries to determine the self-consistent potential landscape and find out which geometry suppresses a local potential dip near the tunnel barrier.